\documentclass[12pt]{iopart}
\usepackage{graphicx}
\usepackage[normalem]{ulem}
\expandafter\let\csname equation*\endcsname\relax
\expandafter\let\csname endequation*\endcsname\relax
\usepackage{amsmath,amssymb}
\usepackage{amsthm}
\newcommand{\ket}[1]{\left\vert#1\right\rangle}
\newcommand{\bra}[1]{\left\langle#1\right\vert}

\begin{document}

\title[Quantum state transfer with ultracold atoms]
{Quantum state transfer with ultracold atoms in optical lattices}

\author{Salvatore Lorenzo}
\address{Quantum Technology Lab, Dipartimento di Fisica, 
Universit\'{a}  degli Studi di Milano, 20133 Milano, Italy}
\address{INFN, Sezione di Milano, I-20133 Milano, Italy}

\author{Tony J. G. Apollaro}
\address{NEST, Istituto Nanoscienze-CNR and Dipartimento di Fisica e Chimica, Universit\'{a} degli Studi di Palermo, via Archirafi 36, I-90123 Palermo, Italy}

\author{Andrea Trombettoni}
\address{CNR-IOM DEMOCRITOS Simulation Center, Via Bonomea 265, I-34136 Trieste, Italy}
\address{SISSA and INFN, Sezione di Trieste, 
Via Bonomea 265, I-34136 Trieste, Italy}

\author{Simone Paganelli}
\address{Dipartimento di Scienze Fisiche e Chimiche, Universit\'{a} dell'Aquila, via Vetoio, I-67010 Coppito-L'Aquila, Italy}

\vspace{10pt}



\begin{abstract}
Ultracold atoms can be used to perform quantum simulations 
of a variety of condensed matter systems, including spin systems. These 
progresses point to the implementation of the manipulation 
of quantum states and to observe and 
exploit the effect of quantum correlations. 
A natural direction along this line is provided by the possibility 
to perform quantum state transfer (QST). After presenting 
a brief discussion of the simulation of quantum spin chains 
with ultracold gases 
and reminding the basic facts 
of QST, we discuss how to potentially 
use the tools of present-day ultracold technology 
to implement the QST between two regions of the atomic system 
(the sender and the receiver). The fidelity and the typical timescale 
of the QST are discussed, together with possible limitations and 
applications of the presented results. 
\end{abstract}

%
%
%
%
%

\section{Introduction}\label{intro}

An important direction of research in the physics of cold atoms and molecules, 
achieved in the last decade, is provided by the possibility to 
prepare in the laboratory 
artificial systems able to mimic more complex condensed matter systems 
\cite{bloch08}. As discussed in the papers of the present JPB Special Issue 
on the implementation of quantum many-body problems with cold atoms and 
molecules, the present technology has made ultracold systems 
extremely accessible with the possibility to address, even locally in limited 
and controllable regions of space, 
most of the system parameters. The combined action of 
trapping potentials, optical lattices and low temperature cooling 
techniques, provide a sort of playground 
to design simulators of a huge class of models. In particular a key ingredient 
in most of the quantum simulations performed with ultracold trapped gases  
is the ability to control the geometry of the system and the capability 
to tailor the one-body external potential felt by the constituents 
of the gas.

These progresses paved the way from one side to the study 
of strongly correlated bosonic and fermionic states, both in the 
continuous space 
and in the presence of optical lattices \cite{lewenstein12,Georgescu2014}, 
and from the other to the control and use of quantum correlations 
\cite{bloch08}. A typical 
and important example in the latter direction is provided by atom 
interferometry with ultracold atoms \cite{pezze16}: 
it has been experimentally showed that using entangled states one can 
have a sensitivity of the atom interferometer below the shot-noise limit 
\cite{gross10,riedel10} and the detection of Bell correlations 
between the spins of $\approx 500$ atoms detected in a Bose-Einstein condensate 
\cite{schmied16} with possible and obvious consequences for the implementation 
of quantum criptography algorithms \cite{nielsenbook}.

Moving forward in the direction of the manipulation and use 
of quantum states, in this paper we address the possibility to perform 
Quantum State Transfer (QST) in ultracold systems. QST has been deeply 
investigated in quantum registers and more generally in spin chains: 
given the fact that ultracold gases can be used to perform a physical 
emulation of spin systems, a natural direction is to investigate 
the implementation of possible QST protocols with cold atoms. Two ingredients 
are instrumental in our study: first, one should use 1D optical lattices 
and realize an effective low-energy spin Hamiltonian \cite{simon11}. Moreover, 
one should be able to identify two regions of sites (say made up by one or two
 sites each) and control the tunneling from the two regions to the 
remaining part of the systems. The two regions are typically referred
as the {\it sender} sites and the {\it receiver} sites. 
The typical lattice unit 
is in the range $0.5-5 \mu m$ \cite{morsch06}: one may 
use optical barriers, which has to be of $\sim 1-2 \mu m$  dimension. 
Controlling such localized potentials, which is experimentally challenging,
one may control the tunneling in the 
direction between the senders and the receivers and the tunneling 
in the other directions (that we argue to be better negligible). 
Such control is now within reach, as demonstrated by the recent experiment 
\cite{valtolina15} in which an optical barrier with an $1/e^2$ 
beam waist at the center of $2.0 \pm 0.2 \mu m$ 
was used to weakly couple two fermionic gases of $^6Li$ atoms, and stable 
enough to observe Josephson oscillations. Finally, one should have the possibility 
to monitor the state of the senders and receivers, which requires the use 
of the techniques of quantum gas microscopy \cite{gross14}.

In this paper we focus our attention on systems composed of 
1D spin chains to be employed as a channel for QST. The structure of the paper 
is the following: After a brief introduction of the possible ways 
we have nowadays to simulate spin lattices by ultracold atoms, 
we give a review on the recent progresses obtained regarding the QST 
of one or two qubits through a spin chain. 
One of the main ingredient of these QST schemes is the partial 
decoupling of  the sender/receiver sites from the rest of the chain  
by strong local magnetic barriers \cite{plastinaPRL2007,lorenzo13,paganelli13,1402-4896-2015-T165-014036,lorenzo15}. In the perspective of a 
possible implementation with ultracold atoms, one possible strategy is to 
avoid local on-site energies. 
To this aim, we propose an alternative scheme for the two-qubit transfer 
that does not requires any local barriers/on-site energies, 
exploiting a weak coupling instead. 
We find that an efficient transfer of the fidelity is still
achievable. Moreover, numerical data show that the transfer times are 
comparable with the one obtained within a similar scheme for the single 
qubit transfer. 
This result could suggest the remarkable fact that the transfer 
rate of a many-qubit  could be only weakly dependent on the number of qubit 
and motivates further analytical studies on this purpose.

\section{Quantum simulation of spin chains}\label{simul}

The study of magnetic systems is a very active field of research 
in solid state physics \cite{blundell01} and correspondingly it developed 
the search of synthetic, controllable physical systems 
having effective magnetic model Hamiltonians 
with tunable geometry and parameters. Ultracold atomic setups provide 
a promising possibility to this aim: 
effective nearest-neighbour interactions for atoms 
in neighbor wells of an optical lattice may result 
from super-exchange couplings and second-order tunneling was been observed 
in superlattices \cite{folling07}.
Classical frustrated magnetism in triangular lattices has been 
simulated with the suitable shaking of optical lattices \cite{struck11}.

A very well studied path to simulate spin chains is to use 
two-component gases: here the two internal degrees of freedom 
emulate the components of the spins. Since spin interactions 
can be tuned rather precisely \cite{bloch08} 
one has an important knob to control the parameters  
of the spin Hamiltonian. Therefore, the 
realization of controllable Bose-Bose mixtures \cite{thalhammer08} 
paves the way towards the experimental simulation of spin Hamiltonians.

Another strategy to simulate antiferromagnetic spin chains was discussed 
in \cite{sachdev02}. The implementation of an Ising chain in a transverse 
field is based on the realization of 
a Mott state in a small applied potential gradient, which exhibits 
a resonant response when the potential energy drop per lattice spacing is 
close to the interatomic interaction energy between two atoms. The presence 
of a small tunneling coupling between neighboring wells turns out to produce 
an effective Ising Hamiltonian \cite{sachdev02}. 
Following the latter suggestion and using a tilted 1D optical lattice,
the Ising chain in a transverse field was 
experimentally implemented \cite{simon11}. 
The detection of the paramagnetic and the antiferromagnetic phases was carried 
out by measuring the probability to have an odd 
occupation of sites \cite{simon11}.
Vibrational mode of trapped ions can also be used to simulate interacting spin systems \cite{Deng2005,Kim2010,giorgi2010,Kim2011,britton2012}.

A simple way to understand how a spin chain Hamiltonian is obtained can 
be seen by considering ultracold bosons in an optical lattice at half-filling 
\cite{giuliano13}: let us consider ultracold bosons in deep optical lattices 
as described by the Bose-Hubbard Hamiltonian \cite{lewenstein12} 
\begin{equation}
H_{\rm BH} = -t\sum_{i=1}^{N-1} 
  \left( b^{\dag}_i b_{i+1} + b^{\dag}_{i+1} b_i \right)+
  \frac{U}{2} \sum_{i=1}^N n_i \left(n_i-1 \right) + 
  V \sum_{i=1}^{N-1} n_i n_{i+1}.
  \label{HAM_GEN}
\end{equation}
In Eq.~(\ref{HAM_GEN}) $b_i$ is the boson operator in the site $i$ 
(with $i=1,\cdots,N$ and $N$ number of sites) and,  
as usual, the parameter $t$ denotes the hopping strength, and $U$ and 
$V$ respectively the interaction 
energy of two particles at the same site and at two nearest neighboring sites 
\cite{baranov12}. Open boundary conditions are used in Eq.~(\ref{HAM_GEN}).

Denoting the filling by $f$, if $f$ is half-integer (e.g., $f=1/2$)  
one can define 
$$s_j^z \equiv n_j-f:$$ 
the eigenvalues of $s_j^z$ are $\pm \frac{1}{2}$. We will later 
use the notation $s_j^\alpha \equiv (1/2) \hat{\sigma}_i^\alpha$ where 
$\alpha=x,y,z$ and $\hat{\sigma}_i^\alpha$ are the Pauli matrices.
For $t=0$ the energy per particle is (for $N \to \infty$) 
$\varepsilon = Uf(f-1)/2+Vf^2$ and therefore one gets, 
in the limit $U \to \infty$,  the effective Hamiltonian
\begin{equation}
  H_{\rm XXZ} = - K \sum_{\langle i,j \rangle} \left( s_i^x s_{j}^x +
  s_i^y s_{j}^y - \Delta s_i^z s_{j}^z \right)
  \label{xxz_GEN},
\end{equation}
where 
\begin{equation}
  K\Delta \equiv V.
  \label{V_0}
\end{equation}
and 
\begin{equation}
  K\equiv 2t \left( f+ \frac{1}{2} \right ).  
  \label{J_0}
\end{equation}
The Glazek-Wilson renormalization group procedure \cite{glazek93} 
(see as well \cite{giuso_1,giuso_2}) provides the XXZ 
Hamiltonian~(\ref{HAM_GEN}) with renormalized coefficients $K$ and $\Delta$ 
\cite{giuliano13}, which are in very goood agreement 
with Bose-Hubbard DMRG numerical results also for $U/t\approx 3$ 
and for small number of sites \cite{giuliano13}. 

These results show than spin chain Hamiltonians emerge naturally for ultracold 
atoms in optical lattices: magnetic fields in the $z$-direction 
are obtained by adding 
on-site energies \cite{morsch06}, while to have transverse fields one may 
resort to schemes like the one discussed in \cite{sachdev02}.   Finally, we 
observe that one can have an XXZ limit by tuning the parameter $\Delta$ 
to zero and that in general a wide class of spin chain models may be obtained. 
In the following, we consider mainly an XX spin chain, but our results, 
with their own peculiarities and modulo analytical and computational 
difficulties, can be extended to XXZ and Ising in a transverse field models \cite{banchi2016}.

\section{Quantum state transfer}\label{review}

In this Section we give a short overview of the class of QST protocols 
without dynamical control over the Hamiltonian and/or 
post-measurement selection procedures. In general, 
the goal of a QST protocol is to transfer the information encoded 
in a quantum state of a system from a sender to a receiver, 
which are located at different spatial positions. The capability of 
faithfully transferring the quantum information between different 
locations by means of a quantum channel plays a key r\^ole 
in many quantum information processing tasks, 
ranging from quantum teleportation to quantum computation and 
cryptography \cite{nielsenbook}. The elementary unit of 
quantum information is referred to as qubit 
and consists of two orthogonal quantum states, 
conventionally labelled by $\ket{0}$ and $\ket{1}$, 
which can belong to a great variety of systems 
where the two-level approximation holds. In the following, we will refer 
to $n$-QST ($n=1,2,...$) depending on the number of qubits 
the protocol aims to transfer. 
Whereas for long-distances 1-QST has been achieved by means of photons 
\cite{Northup2014}, for short-haul transfers it would be more 
advisable to rely on solid-state implementation, in order to avoid, 
e.g., information loss at interfaces between, say, a solid-state based 
quantum processor and a photonic quantum channel. 

Since the seminal proposal by Bose \cite{bose03} 
to use nearest-neighbor coupled spin-$\frac{1}{2}$ 
systems as quantum channels for 1-QST, a great amount of 
work has focused on protocols enhancing the quality of 1-QST. 
Indeed, in the presence of uniform couplings, the quality of 1-QST 
falls below the threshold achievable by means of only local operations 
and classical communication (LOCC) already for spin chains made up 
of a hundred of spins (for a review see, e.g., 
\cite{doi:10.1080/00107510701342313,apollarorev,nikolopoulos2013quantum}). 

Among the protocols without dynamical control, it has been shown
that perfect length-independent 1-QST can be obtained by 
engineering the spin-spin couplings so as to induce a linear dispersion 
relation \cite{Osborne2004,christandl04,difranco08,paganelliFdP2009}, 
which yields a ballistic 1-QST entailing that the time necessary 
to complete the protocol is proportional to the chain length. This result 
has been recently extended also including next-to-nearest neighbour 
couplings \cite{2016arXiv160702639C}. 
A reliable local modulation involving the entire chain, however, 
would face several practical difficulties on the experimental side. 
Ballistic 1-QST can also be achieved under appropriate tuning of the 
outermost couplings \cite{apollaro12, zwick12}.

A different approach relies on the weak interaction of the sender and receiver
spins with a bulk embodied by a uniform chain 
\cite{wojcik05,plenio2005,quantumbus,venuti07-2}. 
Schemes of this kind exploit the appearance of a pair 
of Hamiltonian eigenstates strongly bi-localized at the outermost
weakly-coupled sites, which brings about an effective Rabi-like dynamics 
\cite{wojcik05}. 
A similar dynamics can be triggered by applying strong magnetic fields 
on the sender and receiver qubits or on their nearest neighbors 
\cite{plastinaPRL2007,lorenzo13,paganelli13}.
At variance with ballistic QST protocols, a usual drawback 
of the Rabi-like mechanisms is that they typically require long 1-QST times. 
To this end, in order to reduce the transfer time, proposals superimposing 
a staggered interaction pattern of the couplings~\cite{0295-5075-84-3-30004} 
and a modular partitioning of the staggered quantum channel 
\cite{PhysRevA.93.032310} have been put forward.

Recently, protocols achieving 2-QST have been investigated and it has 
been found that both in the Rabi-like regime, achieved by applying 
strong magnetic fields on the first and last spin 
\cite{1402-4896-2015-T165-014036,lorenzo15} and, 
for the ballistic regime, by engineering all of the inter-spin couplings 
\cite{1367-2630-16-12-123003}, high-quality 2-QST is achievable. 

In this paper we add an original contribution to this topic along this line, 
by showing that the Rabi-like mechanism yielding high-quality 2-QST 
can be performed also by adopting the weak coupling protocol. 
In the following subsection we will briefly illustrate the basic 
mechanism of the 1- and 2-QST protocols.

\subsection{QST in a nutshell}\label{protocol}

\begin{figure}[h]
        \centering
          \includegraphics[width=0.9\textwidth]{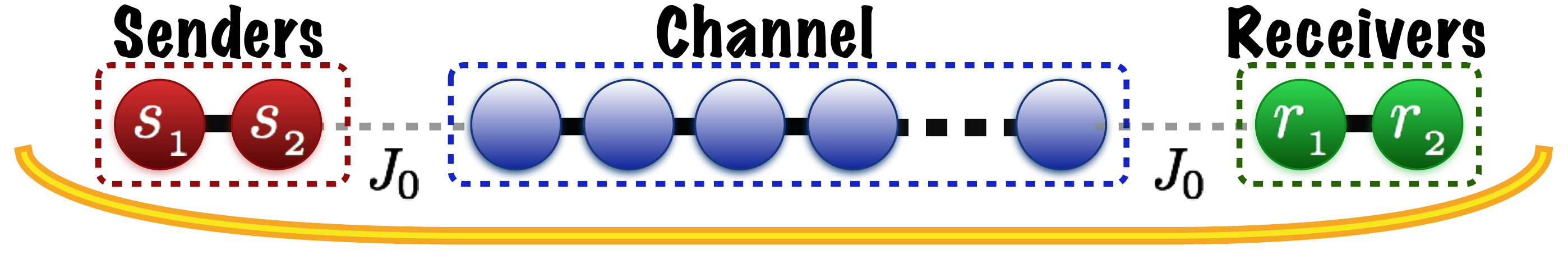}
        \caption{Scheme of n-QST ($n=2$ in this case) depicting the sender and receiver blocks coupled by a quantum channel.
	\label{n-QST}}
      \end{figure}

Following \cite{bose03}, where the 1-QST has been first 
introduced and here generalised to $n$-QST, the sender
prepares an arbitrary state of $n$ qubits, $\ket{\phi}_S$ and
sets the rest of the chain and of the receivers in the fully polarised state 
$\ket{{\mathbf{0}}}_{QC}$ (see Fig.~\ref{n-QST}).
The initial state of the sender block, the quantum channel, 
and the receiver block thus reads 
$|\Psi(0)\rangle=\ket{\phi}_S\ket{{\mathbf{0}}}_{QC}\ket{{\mathbf{0}}}_{R}$.
The system then evolves according to its Hamiltonian $\hat H$ so that, 
at time $t$, its state is given by $\ket{\Psi(t)}= \hat U(t)\ket{\Psi(0)}$ 
with $\hat U(t)= e^{-i \hat H t}$. 
The goal is to exploit such natural dynamics in order to transfer the 
initial sender's state 
$\ket{\phi}_S$ to the qubits embodying the receiver in a given time $\tau $, 
meaning that ideally one aims at obtaining
$\ket{\Psi(\tau )}= \ket{{\mathbf{0}}}_S\ket{{\mathbf{0}}}_{QC}\ket{\phi}_{R}$ 
in, possibly, short times. The state of the receiver is evaluated by  
tracing out the remaining spins, i.e., $\rho_R(\tau)=\mathrm{Tr}_{S,QC}\ket{\Psi(\tau)}\bra{\Psi(\tau)}$, and it is generally a mixed state. 
One thus aims at making the QST fidelity 
$F_\phi(\tau )=_S\!\!\bra{\phi}\rho_R(\tau )\ket{\phi}_S$
as large as possible (the fidelity $F_\phi$ measures how close is the receiver's state to $\ket{\phi}_S$). 
The fidelity introduced above depends on the specific input $\ket{\phi}_S$. 
In order to end up with a state-independent figure of merit for QST protocols,
one averages $F_\phi$ over all possible input states, obtaining thus the 
average fidelity $\bar{F}(t)$.

For Hamiltonians conserving
the total number of excitations one can obtain simple 
expressions for the 1- and 2-QST average fidelity, as $\ket{\Psi(0)}$
is restricted to evolve in the zero- and one-excitation 
subspaces for 1-QST, whereas for 2-QST also the two-excitation sector 
takes part in the dynamics.
The 1-QST average fidelity, for a $N$-site spin system,
is simply given by \cite{bose03}
\begin{equation}\label{E.avFid}
\bar{F}(t)=\frac{1}{2}+\frac{|f (t )|}{3}+\frac{|f (t )|^2}{6}~\,,
\end{equation}
where 
\begin{equation}\label{trans}
f(t)=\langle N|e^{-i \hat H_{\rm ch}t}\ket{1}
\end{equation}
is the excitation transition amplitude from the first to the last spin
(we used the compact notation, referred to as computational basis, 
$\ket{n}\equiv\ket{0}_1\cdots\ket{1}_n \cdots\ket{0}_N$ indicating that 
only the spin residing on site $n$ is flipped to $\ket{1}$). 
Note that $|f (\tau)|=1$ entails $\mathcal F (\tau)=1$ (perfect QST). 
Also, the average fidelity is a monotonic function of the transition 
amplitude and hence the QST performance can be evaluated by just tracking down 
the excitation transport across the array. 
For the 2-QST average fidelity, on the other hand, one obtains 
\begin{eqnarray}\label{E.Fidelitygen}
\bar{F}(t)&=&\frac{1}{4}{+}\frac{5}{54}Re\left[f_{s_1}^{r_1}{+}f_{s_2}^{r_2}+\frac{7}{5}f_{s_2}^{r_2} (f_{s_1}^{r_1})^* +\left(f_{s_1}^{r_1}{+}f_{s_2}^{r_2}\right)(g_{s_1 s_2}^{r_1r_2})^*\right]{+}\frac{1}{54}\left(|f_{s_2}^{r_1}|^2{+}|f_{s_1}^{r_2}|^2 \right)\nonumber \\
&{+}&\frac{5}{108}\left( |f_{s_2}^{r_2}|^2{+}|f_{s_1}^{r_1}|^2 \right){+}\frac{1}{36}|g_{s_1 s_2}^{r_1 r_2}|^2 {+}\frac{7}{54}Re\left[g_{s_1 s_2}^{r_1 r_2}\right]
{-}\frac{1}{54}\sum_{n{=}1}^{n {\not\in}\mathcal{R}}\left(|g_{s_1 s_2}^{nr_1}|^2 + |g_{s_1 s_2}^{nr_2}|^2 \right)
\nonumber\\
&{-}&\frac{1}{27}\sum_{n{=}1}^{n{\not\in}\mathcal{R}}Re\left[(f_{s_2}^{n})^*g_{s_1s_2}^{n r_1}{+}(f_{s_1}^{n})^*g_{s_1 s_2}^{nr_2}\right]~,
\end{eqnarray}
where $f_n^m{=}\bra{m}e^{-i t H_1}\ket{n}$ and
$g_{nm}^{rs}{=}\bra{rs}e^{-i t H_2}\ket{nm}$ are the single- and
two-particle transfer amplitudes from sites $n\rightarrow m$ and
$\{nm\}\rightarrow \{rs\}$, respectively 
\cite{1402-4896-2015-T165-014036,lorenzo15}. 
Finally, if the sender's state is encoded in a qutrit 
(a system with three levels), the average fidelity is reported in 
\cite{Latmiraletal15}.

\section{QST in the $XX$ model}

Motivated by the quantum simulation of spin models with ultracold atoms as 
described in Section \ref{simul}, we focus here on the paradigmatic 
and simpler case of an XX Hamiltonian.

To uniform with the standard notation used on the QST literature, we write 
the $XX$ model of a 1D spin-$\frac{1}{2}$ chain with open boundary conditions 
as
\begin{equation}\label{E.XXHam}
\hat{H}(t)=-\sum_{i=1}^{N-1}J_i\left(\hat{\sigma}_i^x\hat{\sigma}_{i+1}^x+\hat{\sigma}_i^y\hat{\sigma}_{i+1}^y\right)+\sum_{i=1}^{N}h_i \hat{\sigma}_i^z,
\end{equation}
where $\hat{\sigma}^{\alpha}$ ($\alpha=x,y,z$) are the Pauli matrices. 
The Hamiltonian described by Eq.~(\ref{E.XXHam}) conserves the 
total magnetisation in the $z$-direction and hence the formulas reported 
in the previous Section apply. We observe that in typical experimental 
realizations one has an external trapping potential (see 
Fig.~\ref{n-QST}), whose effect is of course 
to increase the density of the center of the trap. However, hard wall 
potentials have been realized \cite{gaunt12} and in general 
the presence of non-shallow boundary conditions is of help 
for the efficiency of QST. In the following we choose the sender and receiver 
sites at the end of the systems: this helps as well the QST efficiency, and 
can be obtained by hard-wall (open) boundary conditions or by the use 
of strong optical barriers, respectively at the left (right) 
of the sender (receiver) sites, as clear from Fig.~\ref{n-QST}. 
As we commented in Section \ref{intro}, 
this requires the capability of controlling laser beams on spatial distances 
of order of few microns. A similar control is needed to vary the coefficients 
$J_i$ at the right (left) 
of the sender (receiver) sites, i.e., in the direction joining sender and 
receivers sites, which in turn appears very useful in increasing the QST 
efficiency.

\subsection{Rabi-like 1-QST}\label{rabi}

In the single-excitation sector, the spectral decomposition 
of Eq.~(\ref{E.XXHam}) reads
$\hat H{=}\sum_{k=1}^N \varepsilon_k  \ket{\varepsilon_k}\!\! \bra{\varepsilon_k}$, where $\varepsilon_k$ is the $k$th energy 
eigenvalue with corresponding eigenstate 
$\ket{\varepsilon_k}{=}\sum_{n=1}^N a_{nk} \ket{n}$ written in the 
computational basis. In this representation, the transition amplitude 
in Eq.~(\ref{trans}) is given by
\begin{equation}\label{E.TranAmp}
f (t)=\sum_{k=1}^N e^{-i  \varepsilon_k t}  a^*_{k N} a_{k 1}=\sum_{k =1}^N e^{-i \varepsilon_k  t} \bra{\varepsilon_k}N\rangle\!\!\bra{1}\varepsilon_k\rangle.
\end{equation}

The last identity shows that each eigenstate contributes to 
Eq.~(\ref{E.TranAmp}) through the quantity 
$\bra{\varepsilon_k}N\rangle\!\!\bra{1}\varepsilon_k\rangle$, 
evolving in time at rate $\varepsilon_k$. 

Various high-quality QST schemes 
\cite{plastinaPRL2007,linneweberetalIJQI2012,lorenzo13,wojcik05,huo08} 
rely on the situation where the edge states $\ket{1}$ and $\ket{N}$ 
have a strong overlap with only two stationary states, say 
those indexed by $j=1,2$ (bi-localization). In this case, 
Eq.~(\ref{E.TranAmp}) can be approximated as  
\begin{equation}\label{f2}
f (t)\simeq  e^{-i \frac{\delta\omega\, t}{2}}a^*_{1N}a_{11}+e^{-i \frac{\delta\omega\, t}{2}}a^*_{2N}a_{21},
\end{equation}
with $\delta \omega{=}\varepsilon_1{-}\varepsilon_2$ (we assumed $\varepsilon_1>\varepsilon_2)$. 
This implies a Rabi-like dynamics that occurs with a characteristic Rabi 
frequency given by $\delta\omega$. Accordingly, $\tau\sim\delta\omega^{-1}$ 
showing that the order of magnitude of the transfer time is set by the 
energy gap between the two bi-localized eigenstates.

The above bi-localization effect is usually achieved 
by introducing perturbation terms in the Hamiltonian that decouple the 
outermost spins from the bulk. This can be realized through: 
{\it (i)} application of strong local magnetic fields on the edge spins 
$n{=}1{,}N$ ~\cite{plastinaPRL2007, linneweberetalIJQI2012}, or 
{\it (ii)} on their nearest-neighbours $n{=}2{,}N{-}1$ \cite{lorenzo13}, 
and {\it (iii)} engineering of weak couplings between 
the edge spins and bulk $J_1{=}J_{N-1}$~\cite{wojcik05,huo08}. 
While all these models share that a pair of Hamiltonian eigenstates 
exhibit strong bi-localizaton on the edge sites, the typical energy gap 
between such two states -- and accordingly the transfer time $\tau$ -- 
depends on the considered protocol. 
Calling $\xi$ the model-dependent perturbation parameter 
(such as the local magnetic field strength),
in {\it (i)} the time scales with $N$ as ${\tau\sim \xi^N}$, 
resulting in a QST time that exponentially increases with the array length, 
whereas in {\it (ii)} and {\it (iii)} the time scales 
as $O(\xi^2)$ and $O(\xi^{-2})$, respectively.

\subsection{Rabi-like 2-QST}\label{rabi2}

If the QST protocol aims at transferring the quantum 
information encoded in two qubits, we need to use 
Eq.~(\ref{E.Fidelitygen}) for the average fidelity 
which involves also transition amplitudes in the two-particle sector. 
A considerable simplification can be achieved by 
taking into account that the Hamiltonian in Eq.~(\ref{E.XXHam}) 
maps into a non-interacting spinless fermion model via the Jordan-Wigner 
transformation and, hence, its dynamics is fully described by the 
single-particle basis. This allows to reduce the two-particle transition 
amplitudes $g_{nm}^{pq}(t)$ to determinants of matrices whose elements 
are single-particle transition amplitudes $f_i^j(t)$, 
where $i{=}\{n,m\}$ and $j{=}\{p,q\}$, e.g.,~\cite{1367-2630-16-12-123003}:
\begin{equation}\label{E.2to1}
g_{nm}^{pq}(t){=}  \begin{vmatrix}

        f_n^p(t) & f_n^q(t) \\

 	f_m^p(t) & f_m^q(t) \\

    \end{vmatrix}.
\end{equation}

In order to achieve an high-fidelity 2-QST via Rabi-like dynamics 
between the sender and the receiver block, each composed of two qubits, 
now we should look for eigenstates that are {\textit{quadri-localized}} 
on the first and last two sites. One way to accomplish it
has been reported in \cite{1402-4896-2015-T165-014036,lorenzo15} 
by means of strong magnetic fields on sites $n{=}3$ and $n{=}N{-}2$. 
In the following, we investigate if also weakly coupling the sender and 
the receiver block to the quantum channel allows for high-quality 2-QST, 
and which are the conditions to be fulfilled for such a result. 

We consider the Hamiltonian (\ref{E.XXHam}) with all the couplings 
uniform but $J_2{=}J_{N{-}2}{=}J_0\ll J \equiv 1$, that is, the sender and 
receiver block are weakly coupled to the quantum channel. As we commented 
in Sec. \ref{simul},, this requires in ultracold systems the possibility to increase the energy 
barrier between on the right (left) of the sender (receiver) blocks, e.g. 
with suitably controlled optical barriers. This scheme is expected to be 
more easily implemented, since the hoppings are be sensibly depending on 
local applied optical barriers 
rather than the on-site energies.

The 2-QST properties of the model described by Eq.~\ref{E.XXHam} 
depend strongly on the number of spins $N$ of the chain, which, 
indeed, determines the spectral properties of the energy eigenvalues.  

By exploiting perturbation theory, justified by the fact that we set $J_0\ll 1$, we can use the result in Ref.~\cite{lorenzo15} for the average fidelity 
obtaining
\begin{eqnarray}\label{E.Fidapp}
&\bar{F_a}(t)=\frac{1}{4}+\frac{10}{54}Re\left[f_1^{N-1}\right]+\frac{7}{54}Re\left[\left(f_1^{N-1}\right)^2\right]+\frac{12}{54}|f_1^{N-1}|^2+\frac{2}{54}|f_1^{N}|^2\nonumber\\
&+\frac{10}{54}|f_1^{N-1}|^2Re\left[f_1^{N-1}\right]-\frac{10}{54}Re\left[f_1^{{N-1}^*}f_1^{N}f_2^{N-1}\right]-\frac{7}{54}Re\left[f_1^{N}f_2^{N-1}\right]~.
\end{eqnarray}

\begin{figure}[htbp]
\includegraphics[width=1\textwidth]{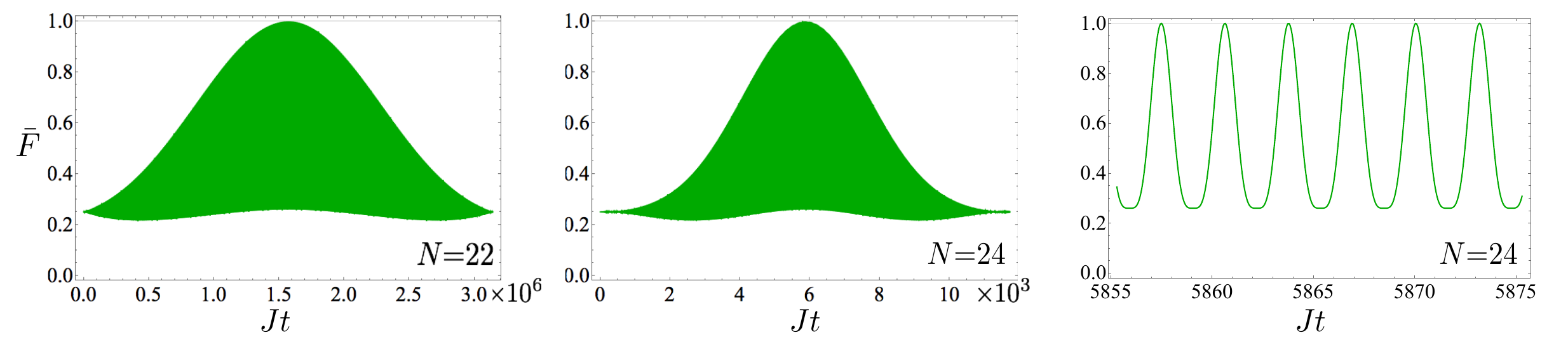}
\caption{Average fidelity $\bar{F}(t)$ as a function of time $t$ for $N{=}22$ and $N{=}24$, shown in the left and middle panel, respectively. In both cases the weak coupling has been set to $J_0{=}0.001$. (right panel) Zoom on the time interval where $\bar{F}(t)$ for $N{=}24$ is maximum (for $N{=}22$ only the time values on the ascissa change accordingly to those reported in the right panel). The oscillations occurring on a time scale proportional to $J$ are due to the coupling between the two spins embodying the senders (receivers). \label{F.time_panel}}\end{figure}

Let us consider the case $h_i{=}0$, for all $n$, in Eq.~(\ref{E.XXHam}). 
For odd $N$, we always have a zero-energy level and, for 
$N{=}6n\pm1$ and $N{=}3\left(2n+1\right)$  ($n{=}1,2,\dots$) 
there are also two sets of double- and triple-degenerate energy levels 
at $E\simeq \pm 2 J$, respectively. In such cases the 2-QST is less 
efficient than the classical transfer, i.e., the 
fidelity attainable by means of LOCC, which reads 
$\bar{F}_{LOCC}{=}2/5$~\cite{PhysRevA.60.1888}, and we will 
not deal with such a case here.
For even $N$, the zero energy level is no longer present and only 
the two sets of double- and triple-degenerate states remain, 
respectively for $N{=}2\left(3n{+}1\right)\vee 2\left(3n{+}2\right)$ 
and $N{=}6n$.
An instance of the 2-QST performance for even $N$ is shown in 
Fig.~\ref{F.time_panel}, where it is evident that, 
although both even-$N$ classes reach high-quality transfer, 
the faster one is given by lengths of the chain fulfilling $N{=}6n$. 
Notice that with $J \sim 1nk$, then $10^{3}Jt$ corresponds to times of seconds, 
which is a long time for ultracold experiments, but not completely irrealistic.
This will be the only case we will analyse in some detail in the following, 
since for $N{\neq} 6n$ the time scale on which the transfer takes place 
is really well beyond experimental accessibility in cold atom gases.

\begin{figure}[htbp]
\includegraphics[width=1\textwidth]{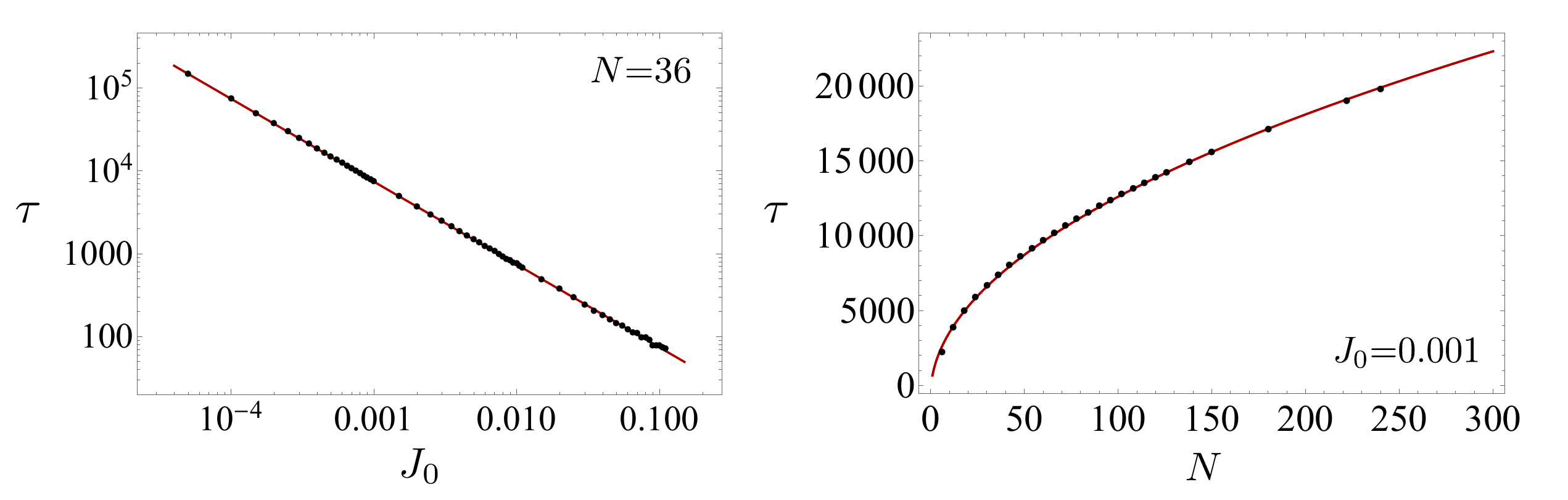}
\caption{
(left panel) Transfer time $\tau$ for which $\bar{F}(\tau)>0.97$ vs. coupling $J_0$ in a $log-log$ plot for fixed values of $N$ yielding  $\tau\simeq J_0^{-1}$ for $J_0\ll 1$. (right panel) Transfer time $\tau$ vs $N$ for a fixed value of $J_0$ (here $J_0{=}0.001$) yielding $\tau\simeq \sqrt{N}$ for $J_0\ll 1$.As a consequence, the transfer time of the 2-QST weak coupling protocol, for $N=6n$, results to be $\tau\simeq \sqrt{N}/J_0$.}\label{F.fits}
\end{figure}

For such a case, we report in Fig.~\ref{F.fits} the transfer time $\tau$, 
with the constraint for the average Fidelity $\bar{F}(t)>0.97$, as a 
function both of the weak coupling value $J_0$ and of the length of the chain. 
An excellent fit of the  transfer time is found by setting 
$\tau \simeq \sqrt{N}/J_0$. 

One can compare this result with  the perturbative analysis of a 
single excitation propagation done in \cite{quantumbus,paganelli13} 
where a single-spin sender is resonantly coupled with the energy band of 
the channel. In the present case the sender is composed by two sites with 
single excitation energies $\epsilon={\pm} 2J$. The single  excitation is 
resonant with the edge of the channel band, while the double excitation is 
resonant with the center of the band. In both cases, it is shown in 
\cite{quantumbus,paganelli13} that the propagation time is also proportional 
to $\sqrt{N}/J_0$.

Following the analysis of \cite{lorenzo15}, we notice that the term 
to be maximized in Eq.~(\ref{E.Fidapp}) is 
$Re\left[ f_1^{N{-}1}\right]$, which can be expressed as
\begin{eqnarray}\label{E.f1N}
Re[f_1^{N{-}1}]=Re[\sum_{k{=}1}^N e^{- i \varepsilon_k t} a_{k1}a_{kN{-}1}] \simeq Re[\sum_{i\in{\mathbf{k}}} e^{- i \varepsilon_{i} t} a_{i 1}a_{i N{-}1}]~,
\end{eqnarray}
where in the last step we have exploited the fact that 
only the two pairs of triple quasi-degenerate levels have a 
significant overlap with the initial state, labelling that set by 
${\mathbf{k}=\{1,2,\dots,6\}}$. By ordering the energy eigenvalues in 
increasing order, one has 
${\mathbf{k}}=\{\frac{N}{3}-1,\frac{N}{3},\frac{N}{3}+1,\frac{2N}{3},\frac{2N}{3}+1,\frac{2N}{3}+2\}$.
Furthermore, exploiting the mirror-symmetry, 
implying that the eigenvectors $\ket{\varepsilon_k}$ are either symmetric 
or antisymmetric \cite{Vaia_tri}, 
and by means of elementary trigonometric identities we obtain
\begin{eqnarray}\label{E.f1N_final}
Re[f_1^{N{-}1}]=\frac{\cos(2 J t)}{2}\left(1+\cos(\delta \omega t)\right)~,
\end{eqnarray}
where $2\delta \omega{=}\varepsilon_{3}{-}\varepsilon_{1}{=}\varepsilon_{6}{-}\varepsilon_{4}$.
The transfer time estimate $\tau=\pi/\delta \omega$ is very accurate, 
giving an error of order $O(1)$, due to the rapid oscillations in the 
prefactor caused by the coupling between the qubits in the sender (receiver) 
block.

On the other hand, for $N{=}2\left(3n{+}1\right)\vee 2\left(3n{+}2\right)$, 
the transfer time $\tau$ does not change significantly as it amounts to 
$\tau\simeq \pi/2\times 10^{-6}$, which has been verified numerically for 
lengths up to a few hundreds. The reason behind this invariance is 
that the energy separation of the doublet of degenerate energy levels 
pair is a secord order effect in $J_0$.

\section{Conclusions}\label{Concl}

In this paper we presented an overview on the potential application of 
ultracold atomic systems in the context of quantum state transfer (QST). 
Among the many-body quantum systems 
one can mimic with ultracold atoms in optical lattices, 
there is a wide class of spin-$1/2$ chains which in turn provide 
the building block of a vaste number of protocols proposed for QST. 
Under a suitable initialization, 
a spin chain can be used as a quantum channel  for a quantum state, 
encoded on a local sender, to be transmitted to a receiver located 
at a different spatial position. 

We briefly reviewed different techniques 
used to simulate quantum spin chains by ultracold atoms. 
Then, we presented the state-of-art on the  QST protocols, 
focusing on those which do not require dynamical control over the 
interactions. We discussed protocols based on $XX$ spin chains, 
involving strong local magnetic fields between sender/receiver 
and the spin chain.  An almost perfect state transfer 
can be achieved, both for a single qubit and a two-qubit state and, 
contrary to a ballistic evolution, the presence of localized states 
in the spectrum determines the emergence of a Rabi-like evolution 
yielding very high QST-fidelity. 

As an alternative to the strong local magnetic fields, 
whose threshold value for efficient QST could be experimentally 
difficult to implement, we studied a similar protocol 
based on a weak coupling between sender/receiver and the channel. 
We have shown that also in this case a Rabi-like process dominates 
the dynamics and an efficient QST can occur also for two-qubit states. 
Even if the problem of  2-QST is formally more complex than the single 
qubit case, nevertheless we found that 
the transmission times are of the same order of magnitude. 
Numerical results suggest that the entire dynamics is dominated by 
an evolution similar to that of a single-excitation and, 
as a consequence, the introduction of an extra qubit does not affect very 
much the transmission efficiency with respect to the single-qubit case.
Also within this protocol,  eigenstates localized both on the sender 
and receivers can be found for chains composed by an even number of spins. 
This allows us to simplify the average fidelity to give a very 
accurate estimation of the transfer times. 
A drawback of this approach resides in the  very large transmission times required, nevertheless, we have shown that for a subclass of even-length 
chains these times can be broadly 
compatible with possible experimental times. 

This paper provide a step towards further studies of QST in 
ultracold setups. An interesting future work would be from one side 
the determination of an effective Hamiltonian description 
for these class of protocols. This would simplify the large-$N$ analysis 
and provide a simple model as a tool to optimize and simulate 
different protocols. A systematic comparison between QST in different 
spin chains (like the XXZ) would be also desirable, and in particular to 
compare the QST efficiency in these models of protocols based on the control 
of the couplings and of the local magnetic fields.

In order to better adhere to the experimental settings,
one should consider the presence of a site-dependent 
local energy in the Hamiltonian, due to the trapping potential 
the system is confined in. Although much of the analysis presented 
in this review, including the formal expressions for the dynamics and 
the fidelity, still hold, new interference term would appear which 
could have a detrimental effect on the QST performance.

It would also be interesting to consider QST protocols where the sender and the receiver are not located at the edges of the 1D system but in the bulk. 
Indeed, due to the natural inhomogeneity of the trap, the edge sites are 
liable to static imperfections, as well as during the initial state 
preparation. As we discussed, a possibility to avoid this 
drawback is to isolate the sender and the receiver inside the bulk 
of the spin chain by effectively decoupling them from the edges by means of 
two extra optical barriers.
Due to the achievable site distance and compared with typical (stable) laser 
widths, this should be challenging but 
feasible and it would interesting to perform a 
more detailed analysis along this line in connection with experimental 
progresses.

\section{Acknowledgments} \label{ack}

Discussions with F. Buccheri, D. Giuliano, T. Macr\`i 
and P. Sodano are gratefully acknowledged. S.P. is supported by a Rita Levi-Montalcini fellowship of MIUR.
T.J.G.A. and S.L. acknowledge funding under the EU Collaborative 
Project TherMiQ (Grant No. 618074) and the UE grant  QuPRoCs 
(Grant Agreement 641277).
\vspace{10pt}

\bibliography{QST.bib}

\end{document}